\documentclass[9pt,twocolumn,twoside]{pnas-new}

\usepackage{xr}
\templatetype{pnasresearcharticle}

\title{How social information can improve estimation accuracy in human groups}

\author[a,b]{Bertrand Jayles}
\author[c]{Hye-rin Kim}
\author[b]{Ram\'on Escobedo}
\author[d]{St\'ephane Cezera}
\author[e,f]{Adrien Blanchet}
\author[g]{Tatsuya Kameda}
\author[a]{Cl\'ement Sire}
\author[b,e,1]{Guy Theraulaz}

\affil[a]{Laboratoire de Physique Th\'eorique, Centre National de la Recherche Scientifique (CNRS), Universit\'e de Toulouse (Paul Sabatier), 31062 Toulouse, France}
\affil[b]{Centre de Recherches sur la Cognition Animale, Centre de Biologie Int\'egrative (CBI), CNRS, Universit\'e de Toulouse (UPS), 31062 Toulouse, France}
\affil[c]{Department of Behavioral Science, Hokkaido University, Kita 8-jo Nishi 11-chome Kita-ku, Sapporo, Hokkaido, Japan}
\affil[d]{Toulouse School of Economics, INRA, Universit\'e de Toulouse (Capitole), 31000 Toulouse, France}
\affil[e]{Institute for Advanced Study in Toulouse, 31015 Toulouse, France}
\affil[f]{Toulouse School of Economics, Universit\'e de Toulouse (Capitole), 31000 Toulouse, France}
\affil[g]{Department of Social Psychology, The University of Tokyo, 7-3-1 Hongo, Bunkyo-ku, Tokyo, Japan 113-0033}

\leadauthor{Jayles}

\significancestatement{
Digital technologies deeply impact the way people interact.
Therefore, it is crucial to understand how social influence affects
individual and collective decision-making. We performed
experiments where subjects had to answer questions and then
revise their opinion after knowing the average opinion of some
previous participants. Moreover, unbeknownst to the subjects,
we added a controlled number of virtual participants always
giving the true answer, thus precisely controlling social information. Our
experiments and data-driven model show how social influence
can help a group of individuals collectively improve its performance
and accuracy in estimation tasks, depending on the
quality and quantity of information provided. Our
model also shows how giving slightly incorrect information could
drive the group to a better performance.
}

\authorcontributions{B.J., C.S., G.T. designed research; B.J., H.K., R.E., S.C., A.B., T.K., C.S., G.T. performed research; B.J., C.S., G.T. analyzed data; B.J., C.S., G.T. wrote the paper, and T.K. provided critical insights.}
\authordeclaration{The authors declare no conflict of interest.}
\correspondingauthor{\textsuperscript{1}To whom correspondence should be addressed. E-mail: guy.theraulaz{@}univ-tlse3.fr}

\keywords{social influence $|$ wisdom of crowds $|$ collective intelligence $|$ self-organization $|$ computational modelling}

\begin{abstract}
In our digital and connected societies, the development of social networks, online shopping, and reputation systems raises the question of how individuals use social information, and how it affects their decisions.  We report experiments performed in France and Japan, in which subjects could update their estimates after having received information from other subjects. We measure and model the impact of this social information at individual and collective scales. We observe and justify that when individuals have little prior knowledge about a quantity, the distribution of the logarithm of their estimates is close to a Cauchy distribution. We find that social influence helps the group improve its properly defined collective accuracy. We quantify the improvement of the group estimation when additional controlled and reliable information is provided, unbeknownst to the subjects.
We show that subjects' sensitivity to social influence permits to define five robust personality traits and increases with the difference between personal and group estimates.
We then use our data to build and calibrate a model of collective estimation, to analyze the impact on the group performance of the quantity and quality of information received by individuals. The model quantitatively reproduces the distributions of estimates and the improvement of collective performance and accuracy observed in our experiments. Finally, our model predicts that providing a moderate amount of incorrect information to individuals can counterbalance the human cognitive bias to systematically underestimate quantities, and thereby improve collective performance.
\end{abstract}

\doi{\url{www.pnas.org/cgi/doi/10.1073/pnas.XXXXXXXXXX}}

\begin{document}
\maketitle

\ifthenelse{\boolean{shortarticle}}{\ifthenelse{\boolean{singlecolumn}}{\abscontentformatted}{\abscontent}}{}

\dropcap{I}n a globalized, connected, and data-driven world, people rely increasingly on online services to fulfill their needs. AirBnB, Amazon, Ebay, or Trip Advisor,  to name just a few, have in common the use of feedback and reputation mechanisms \cite{dellarocas_2003} to rate their products, services, sellers, and customers. Ideas and opinions increasingly propagate through social networks such as Facebook or Twitter  \cite{cha_measuring_2010,jansen_twitter_2009,goncalves_social_2015}, to the point that they have the power to cause political shifts \cite{bond_61-million-person_2012}. In this context, it is crucial to understand how social influence affects individual decision-making, and its resulting effects at the  level of a group.

Two observations can be made about these collective phenomena: (a) people often make decisions not simultaneously but sequentially \cite{bikhchandani_1992,banerjee_1992}, and (b) decision tasks involve judgmental/subjective aspects. Social psychological research on group decision-making has established that consensual processes vary greatly depending on the demonstrability of answers \cite{laughlin_2011}. When the solution is easy to demonstrate, people often follow the ‘‘truth-wins’’ process, whereas when the demonstrability is low, they are much more susceptible to ‘‘majoritarian’’ social influence \cite{kameda_cognition}. Thus, collective estimation tasks where correct solutions cannot be easily demonstrated are particularly well suited for measuring the impact of social influence on individuals' decisions. Galton’s original work  \cite{levy_galtons_2002} on estimation tasks shows that the mean or median of \emph{independent} estimates of a quantity can be impressively close to its true value. This phenomenon has been popularized as the Wisdom of Crowds (WOC) effect \cite{surowiecki_wisdom_2005} and is generally used to measure a group’s performance. Yet, because of the independence condition, it does not consider potential effects of social influence.

In recent years, it has been debated whether social influence is detrimental to the WOC or not:
some works argue that it reduces group diversity without improving the collective error
 \cite{lorenz_how_2011,mavrodiev_quantifying_2013},
while others show that it is beneficial, if one defines collective performance otherwise
\cite{kerckhove_modelling_2015,luo_social_2016}. One or two of the following measures were
used to define performance and diversity: let us define $E_i$ as the estimate of individual $i$, $\langle E_i\rangle$ as its average over all individuals, and $T$ as the true value of the quantity to estimate.
Then, $\mathcal{G}_D = \langle (E_i - \langle E_i \rangle)^2 \rangle$ is a measure of group diversity, and
$\mathcal{G} = (\langle E_i \rangle - T)^2$ and $\mathcal{G}' = \langle (E_i - T)^2 \rangle$
are two natural measures of the group performance. However, these estimators are not independent, since $\mathcal{G}' = \mathcal{G} + \mathcal{G}_D$, which shows that a decrease in diversity $\mathcal{G}_D$ is beneficial to group performance, as measured by $\mathcal{G}'$, contrary to the general claim.
Later research showed that social influence helps the group perform better, if one considers only information coming from informed \cite{faria_leadership_2010}, successful \cite{king_is_2012}, or confident \cite{madirolas_improving_2015} individuals. We will show that these traits are actually strongly related.
The way social information is defined also matters: providing individuals with the arithmetic or geometric mean of estimates of other individuals has different consequences \cite{madirolas_improving_2015}.

Besides these methodological issues, it is difficult to precisely analyze and characterize the impact of social influence on individual estimates without controlling the quality and quantity of information that is exchanged between subjects. Indeed, human groups are often composed of individuals with heterogeneous expertise, so that in a collective estimation task, one cannot rigorously control the quality and quantity of shared social information, and the quantification of individual sensitivity to this information is hence very delicate. To overcome this problem, we performed experiments in which subjects were asked to estimate quantities about which they had very little prior knowledge (low demonstrability of answers), before and after having received social information. The interactions between subjects were sequential and local, while most previous works have used a global kind of interaction, all individuals being provided some information (estimates of other individuals in the group) at the same time \cite{madirolas_improving_2015,lorenz_how_2011,mavrodiev_quantifying_2013,yaniv_receiving_2004,
kerckhove_modelling_2015}. From the individuals' estimates and the social information they received, we were able to deduce their sensitivity to social influence. Moreover, by introducing \emph{virtual experts} (artificial subjects providing the true answer, thus affecting social information) in the sequence of estimates -- without the subjects being aware of it -- we were able to control the \emph{quantity and quality of information} provided to the subjects, and to quantify the impact of this information on the group performance.

Our results show that the subjects' reaction to social influence is heterogeneous and depends on the distance between personal and group opinion.
We then use the data to build and calibrate a model of collective estimation, to analyze and predict the impact of information quantity and quality received by individuals on the performances at the group level.

\section*{Experimental design}

Subjects were asked to answer questions in which they had to estimate various social, geographical, astronomical quantities, or the number or length of objects in a picture. For each question, the experiment proceeded in two steps: subjects had first to provide their personal estimate $E_p$. Then, after receiving the social information $I$, they were asked to give a new estimate $E$. $I$ is defined as the geometric mean of the $\tau$ previous estimates $E$ ($\tau = 1$ or 3). Subjects answered each question sequentially (see SI Appendix, Fig.~S1), and were not told the value of $\tau$. Since humans think in terms of orders of magnitude \cite{dehaene_log_2008}, we used the geometric mean for $I$ -- which averages orders of magnitude -- rather than the arithmetic one.

Virtual ‘‘experts’’ providing the true value $E=T$ for each question were inserted at random into the sequence of participants (see SI Appendix, Fig.~S1). For each sequence involving 20 human participants, we controlled the number $n = 0$, 5, 15, 80, and hence the percentage $\rho = \frac{n}{n+20} =  0\,\%$, $20\,\%$, $43\,\%$ or $80\,\%$ of virtual experts. The social information delivered to human participants, being the geometric mean of previous estimates, is hence strongly affected by these virtual experts.

When providing their estimates $E_p$ and $E$, subjects had to report their confidence level in their answer on a Likert scale ranging from $1$ (very low) to $5$ (very high), and were asked to choose the reason that best explained their second estimate among a list of 8 possibilities. We used initial conditions for the social information $I$ chosen reasonably far from the true answer $T$, and imposed loose limits to the estimates subjects could give, to prevent them from answering too absurdly. All graphs presented here are based on the 29 questions ($5394\times 2$ prior and final estimates) from the experiment performed in France. A similar experiment was conducted in Japan, for which all results can be found in SI Appendix, where the full experimental protocol is described in detail.

The aims and procedures of the experiments conformed to the ethical rules imposed by the Toulouse School of Economics  and the Center for Experimental Research in Social Sciences at Hokkaido University. All subjects in France and Japan provided written consent for their participation.

\section*{Results}

\subsection*{Distribution of individual estimates} \label{dist_estim}

Previous works have shown that distributions of independent individual estimates are generally highly right-skewed, while distributions of their common logarithm are much more symmetric \cite{madirolas_improving_2015,lorenz_how_2011,mavrodiev_quantifying_2013}. This is because humans think in terms of orders of magnitude, especially when large quantities are involved, which makes the logarithmic scale more natural to represent human estimates \cite{dehaene_log_2008}. In these works, participants were mostly asked ``easy'' questions for which they had good prior knowledge (high demonstrability), such that the answers ranged over 1 to 2 orders of magnitude at most \cite{lorenz_how_2011,madirolas_improving_2015,moussaid_social_2013,mavrodiev_quantifying_2013,king_is_2012,
harries_combining_2004,chacoma_opinion_2015,yaniv_receiving_2004,kerckhove_modelling_2015}. In order to ensure that little information was present before the inclusion of our virtual experts, and to more clearly identify the impact of social influence, we selected ``hard'' questions (low demonstrability). These questions involve very large quantities and answers span several orders of magnitude, making the log-transform of estimates even more relevant. To compare quantities that can differ by orders of magnitude, we normalize each estimate $E$ by the true answer $T$ to the question at hand, and define the log-transformed estimate $X = \log(\frac{E}{T})$. Note that the log-transform of the actual answer $T$ is $X=0$.

\begin{figure}
\centering
	\includegraphics[width=0.48\textwidth]{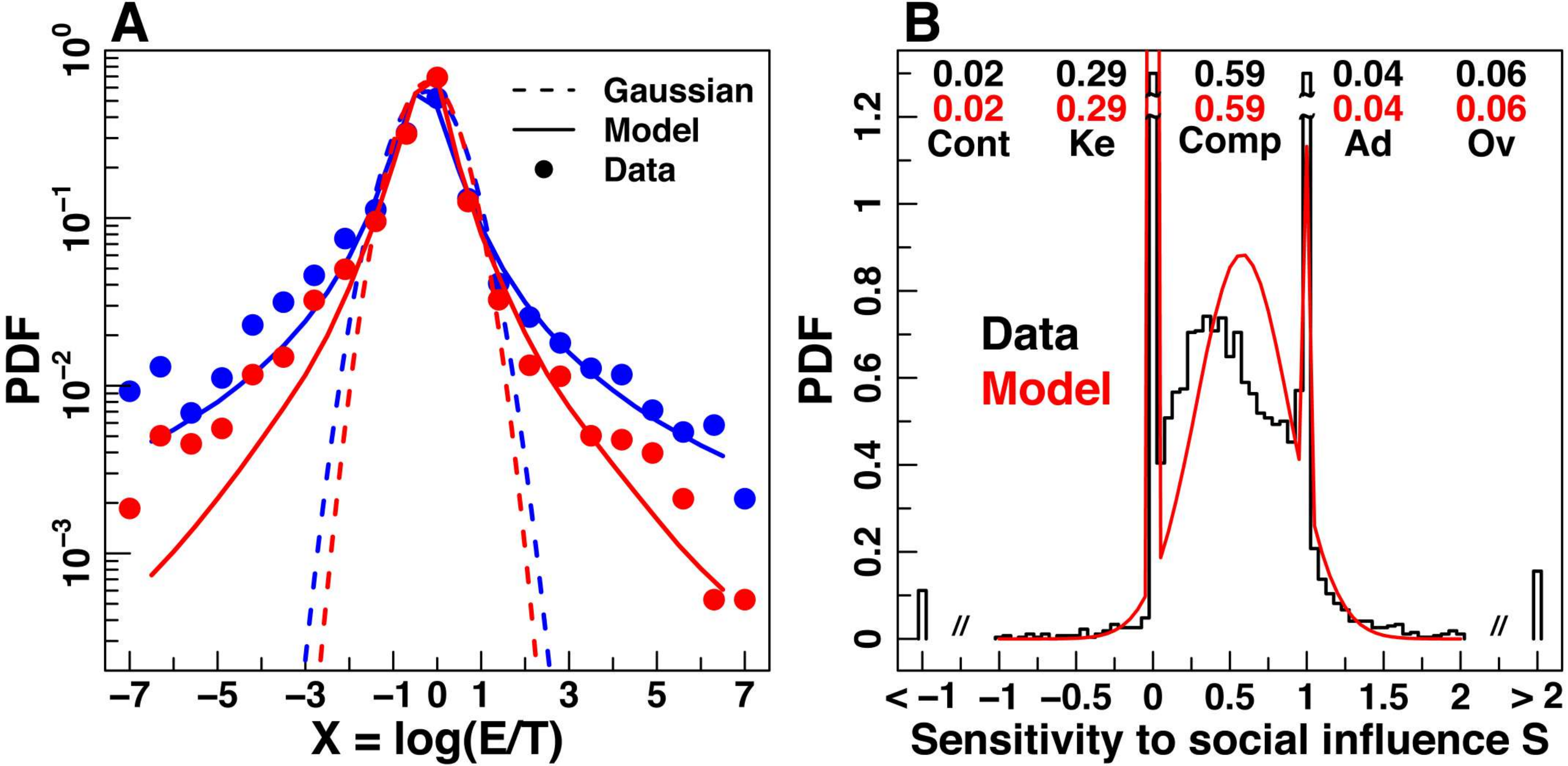}
		\caption{A. Probability  distribution function (PDF) of log-transformed normalized estimates $X = \log(\frac{E}{T})$, where $E$ is the subject's estimate and $T$ is the true answer to the question, before (blue) and after (red) social influence. All conditions ($\rho,\tau$) are aggregated (see SI Appendix, Fig.~S3 for each $\rho$). Plain lines are the results of our model based on Cauchy distributions, while dotted lines are Gaussian fits; B. PDF of sensitivities to social influence $S$. The numbers at the top of the panel are the probabilities for each category of behavior: contradict (Cont: $S<0$), keep (Ke: $S=0$), compromise (Comp: $0<S<1$), adopt (Ad: $S=1$) and overreact (Ov: $S>1$). Experimental data are shown in black, and numerical simulations of the model are in red. The full range of $S$ goes from $-30$ to $200$. The figure is limited to the interval [-1, 2] and the values of S outside this range were grouped in the boxes $S < -1$ and $S > 2$.}
		\label{fig1}
\end{figure}

Figure~\ref{fig1}A shows the distribution of $X$ before and after social information has been provided to the subjects (see also SI Appendix, Table~S1). Although such distributions have often been presented as close to Gaussian distributions \cite{madirolas_improving_2015,mavrodiev_quantifying_2013}, we find that they are much better described by Cauchy distributions, because of their fat tails which account for the non negligible probability of estimates extremely far from the truth. The Cauchy probability distribution function reads
\begin{align}
	f(X,m,\sigma) = \frac{1}{\pi} \frac{\sigma}{(X-m)^2 + \sigma^2},
\end{align}	
where $m$ is the center/median and $\sigma$ is the width of the distribution. SI Appendix, Fig.~S2A shows the distribution of estimates in the Japan experiment, and SI Appendix, Fig.~S2B shows that, when the same questions were asked, distributions of personal estimates in France and Japan are almost identical.

For the Cauchy distribution, the mean and standard deviation are not defined. Therefore, good estimators of $m$ and $\sigma$ are respectively the median and half the interquartile range (the difference between the third and first quartile) of the experimental distribution. In the following, $m_p$ (respectively $m$) and $\sigma_p$ (respectively $\sigma$) will refer to the median and half the interquartile range of the experimental distribution before social influence (respectively after social influence).

Cauchy and Gaussian distributions belong to the so-called stable distributions family.
More generally, $\{X_i\}$ being a set of estimates drawn from a symmetric probability distribution $f$ characterized by its center $m$ and width $\sigma$, we define the weighted average $X' = \sum_i p_i X_i $, with $\sum_i p_i = 1$. $f$ is a stable distribution if $X'$ has the \emph{same} probability distribution $f$ as the original $X_i$, up to the new width $\sigma'$. Indeed, the center $m$ remains the same due to the condition $\sum_i p_i = 1$, but the width may decrease after averaging (law of large numbers), depending on the stable distribution $f$ considered.
Cauchy and Gaussian represent two extremes of the stable distribution family, L\'evy distributions being intermediate cases: for the Cauchy distribution, the width $\sigma$ remains \emph{unchanged}, whereas the narrowing of $\sigma$ is \emph{maximum} for the Gaussian distribution (see SI Appendix). In the case of actual human estimates, the relevance of a certain distribution $f$ can be related to the degree of \emph{prior knowledge} of the group. When individuals have no idea about the answer to a question, the weighted average of arbitrary answers cannot be statistically better ($\sigma'<\sigma$) or worse ($\sigma'>\sigma$) than the arbitrary answers themselves, leading to a Cauchy distribution for these estimates (the \emph{only} distribution for which $\sigma'=\sigma$). However, when there is a good prior knowledge, one expects that combining answers gives a better statistical estimate ($\sigma'<\sigma$; Gaussian). When the quantity to estimate is closely related to general intuition (ages, dates...), estimates should hence follow a Gaussian-like distribution, while when individuals have very little knowledge about the answer, as in our experiment, estimates should be Cauchy-like distributed. The rationale for naturally observing stable distributions is explained in SI Appendix.

We use the term Cauchy-like because Figure~\ref{fig1}A shows that the distributions of prior ($X_p$) and final ($X$) estimates are slightly skewed toward low estimates ($X<0$), reminiscent of the human cognitive bias to underestimate numbers, due to the nonlinear internal representation of quantities \cite{indow_scaling_1977}. As we will show, this phenomenon has strong implications on the influence of information provided to the group. We also observe a clear sharpening of the distribution of estimates after social influence, mainly due to the presence of the virtual experts, hence affecting the value of the social information $M=\log(\frac{I}{T})$, and ultimately, the final estimate $X$ of the actual subjects. This sharpening becomes stronger as the percentage of experts increases (SI Appendix, Fig.~S3).

Moreover, consistently with our introductory discussion of the measurement methods of group performance, we propose the two following indicators: (1) \emph{collective performance}: $|{\rm median}(X_i)|$, which represents how close the center of the distribution is to 0 (the log-transform of the true value $T$), and (2) \emph{collective accuracy}: ${\rm median}(|X_i|)$, which is a measure of the proximity of individual estimates to the true value.

\subsection*{Distribution of individual sensitivities to social influence} \label{distS}

After having received social information, an individual $i$ may reconsider her personal estimate ${E_p}_i$. The natural way for humans to aggregate estimates is to use the median \cite{harries_combining_2004} or the geometric mean \cite{madirolas_improving_2015}, which both tend to reduce the effect of outliers. Here, the social information we provided to the subject was the geometric mean of the $\tau$ previous answers (including that of the virtual experts providing the true answer $E_i =T$), $I_i = (\prod_{j=i-\tau}^{i-1} E_j)^{\frac{1}{\tau}}$.
Moreover, one can always represent the new estimate $E_i$ as the weighted geometric average of the personal estimate ${E_p}_i$ and the social information $I_i$. Hence, we can uniquely define the \emph{sensitivity to social influence} $S_i$, by $E_i = {{E_p}_i}^{1-S_i} \, {I_i}^{S_i}$. $S_i=0$ corresponds to subjects keeping their initial estimates, while $S_i=1$ corresponds to subjects adopting the estimate of their peers. In terms of log-transformed variables $X_i=\log(\frac{E_i}{T})$, we obtain
\begin{align} \label{main_eq}
	X_i = (1-S_i) {X_p}_i + S_i M_i,
\end{align}	
where the log-transformed social information is simply the arithmetic mean $M_i = \frac{1}{\tau}
{\sum}_{j=i-\tau}^{i-1} X_j$,
and thus $S_i = \frac{X_i - {X_p}_i}{M_i - {X_p}_i}$. Note that in this language, $S_i$ is simply the barycenter coordinate of the final estimate in terms of the initial personal estimate and the social information.

Figure~\ref{fig1}B shows that the experimental distribution of $S$ has a bell-shaped part, that we roughly assimilate to a Gaussian, with two additional Dirac peaks exactly at $S = 0$ and $S=1$ (see SI Appendix, Table~S2 for numerical values). Five types of behavioral responses can be identified: keeping one’s opinion (peak at $S = 0$), adopting the group’s opinion (peak at $S = 1$), making a compromise between one’s opinion and the group's opinion ($0 < S < 1$), overreacting to social information ($S > 1$), and contradicting it ($S < 0$). Quite surprisingly, responses that consist in overreacting and contradicting are generally overlooked in previous works \cite{chacoma_opinion_2015,moussaid_social_2013,soll_strategies_2009,harries_combining_2004}, either considered as noise and simply not taken into account, or sometimes included into the peaks at $S=0$ and $S=1$, despite these behaviors being not negligible (especially overreacting).
We find that the median of $S$ is $0.34$, in agreement with previous results \cite{madirolas_improving_2015, soll_strategies_2009,luo_social_2016}, meaning that individuals tend to give more weight to their own opinion than to information coming from others \cite{yaniv_receiving_2004,kerckhove_modelling_2015}.
Moreover, the distribution of $S$ for the experiment performed in Japan and for men and women (in France) are very similar to that of Figure~\ref{fig1}B (see SI Appendix, Fig.~S4).

We find that the subjects' behavioral reactions are highly consistent, reflecting robust differences in personality or general knowledge: in each session, according to the way subjects modified their estimates on average in the first $24$ questions, we split the subjects into three subgroups. We first define \textit{‘‘confident’’} subjects as the quarter of the group minimizing $\langle|S_q|\rangle_q$, where $q$ is the index of the questions (\emph{i.e.} the subjects who were on average closest to $S = 0$) and the \textit{‘‘followers’’} as the quarter of the group minimizing $\langle|1 - S_q|\rangle_q$ (\emph{i.e.} closest to $S = 1$). The other half of the group is defined as the {\it‘‘average’’} subjects. SI Appendix, Fig.~S5 shows the distributions of $S$ for the three subgroups, computed from questions 25 to 29. The differences are striking (see also SI Appendix, Fig.~S6): for the group of confident subjects, the peak at $S=0$ is about $7$ times higher than the peak at $S=1$, while for the group of followers, it is less than twice larger. Moreover, the distribution for average subjects is found to be very close to the global distribution, shown in Figure~\ref{fig1}B.

\subsection*{Impact of the difference between personal and group's opinions on individual sensitivity to social influence}

\begin{figure}[ht]
		\centering
		\includegraphics[width=.48\textwidth]{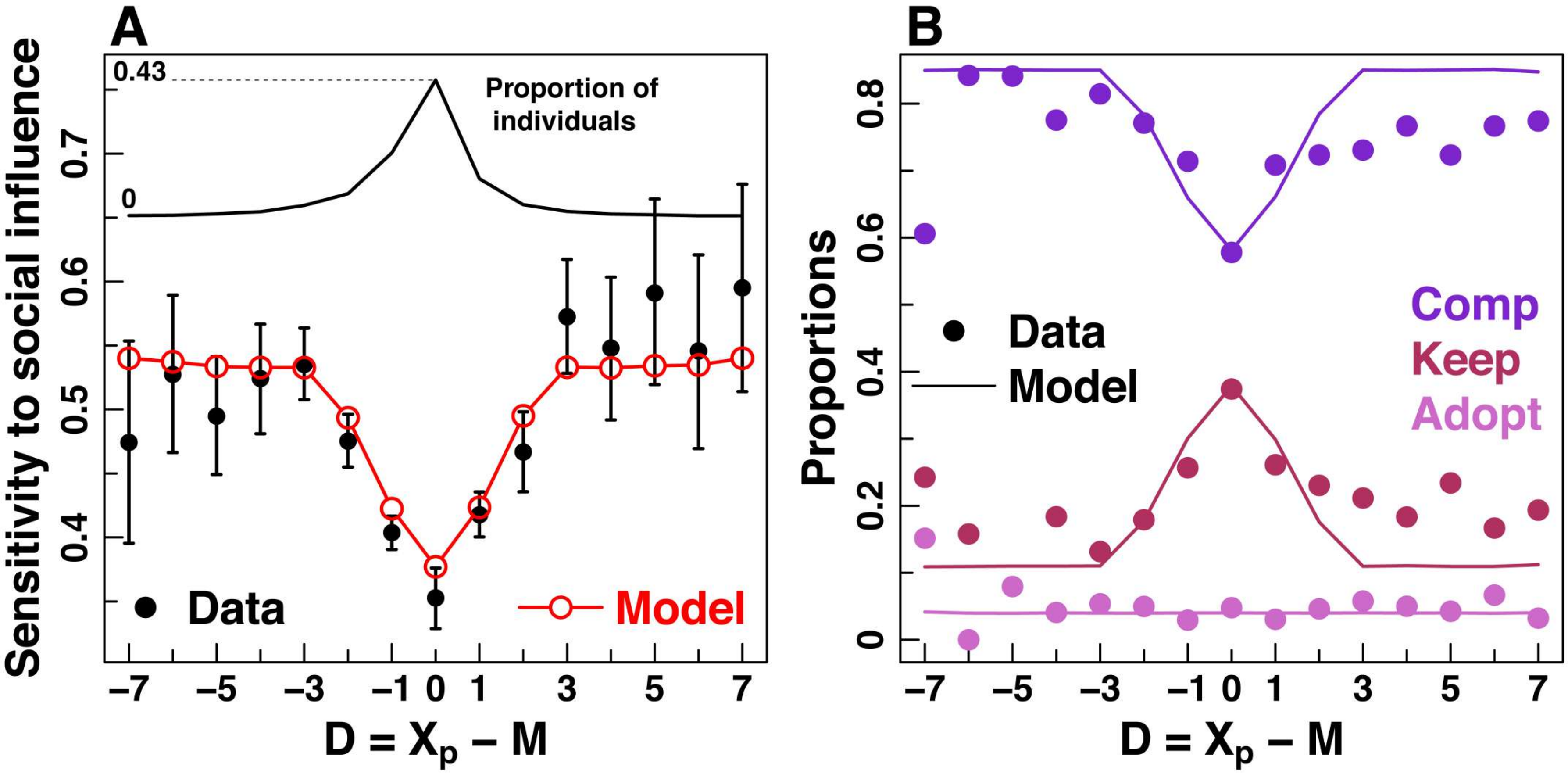}
		\caption{A. Mean sensitivity to social influence $S$ against the distance $D = X_p - M$ between personal estimate $X_p$ and social information $M$ (group estimate). Full black circles correspond to experimental data, while red empty circles are simulations of the model. Note that beyond $3$ orders of magnitude are only about $14\,\%$ of data; B. Fraction of subjects keeping (maroon), adopting (pink), and being in the Gaussian-like part of the distribution of $S$ (mostly compromisers; purple) against $D$.}
		\label{fig4}
\end{figure} 	

Figure~\ref{fig4}A shows that, on average, $S$ depends on the distance $D = \log(\frac{E_p}{I}) = X_p - M$, between personal and group estimates. Up to a threshold of $t \approx 2.5$ orders of magnitude, there is a linear cusp relation between $S$ and $D$.
The farther away the social information $M$ is from a subject's personal estimate $X_p$, the more likely the latter is to trust the group, as $S$ increases. Figure~\ref{fig4}B shows the origin of this correlation: as social information gets farther from personal opinion, the probability to keep one's opinion ($S=0$) decreases, while the probability to compromise increases. Interestingly, the adopting behavior does not change with $D$.
The same phenomena have been observed in the Japan experiment (SI Appendix, Fig.~S8).

\subsection*{Model}

 We now introduce an individual-based model to understand the respective effects of individual sensitivity to social influence and information quality and quantity on collective performance and accuracy observed at the group level. In the model, we simulate a sequence of $20$ successive estimates performed by the agents (not including the virtual experts). A typical run of the model consists of the following steps, for a given condition $(\rho,\tau)$:

1. An initial condition $X_0$ is chosen at random, according to the experimental ratios of initial conditions;

2. With probability $\rho$, the true value $0$ is introduced into the sequence, and with probability $(1-\rho)$, an agent plays;

3. The agent first determines its personal estimate $X_p$ from a Cauchy distribution $f(X_p,m_p,\sigma_p)$, restricted to $[-7;7]$;

4. The agent receives, as social information, the average of the $\tau$ previous final estimates $M$;

5. The agent chooses its sensitivity to social influence $S$ consistently with the results of Figure~\ref{fig1}B and \ref{fig4}. In particular, $S$ is drawn in a Gaussian distribution of mean $m_{\rm g}$ with probability $P_{\rm g}$, or takes the value $S=0$ or $S=1$ with probability $P_0$ and $P_1=1-P_0-P_{\rm g}$. $P_0$ and $P_{\rm g}$ have a linear cusp dependence with $D=X_p-M$, while $P_1$ is kept independent of $D$.
For a given value of $D$, the average sensitivity is $\langle S \rangle = P_0 \times 0 + P_1 \times 1 + P_{\rm g} \times m_{\rm g} = \alpha + \beta |D|$, where $\alpha$ and the slope $\beta$ are extracted from Figure~\ref{fig4}A. $P_{\rm g}$ is hence given by $P_{\rm g} =(\alpha + \beta |D| - P_1)/m_{\rm g}$.
The threshold $t$ is determined consistently by the condition  $S_{\rm max} = \alpha + \beta t$, where $S_{\rm max}$ is the value of the plateau beyond $t$ in Figure~\ref{fig4}A. The values of all parameters are reported in SI Appendix, Table~S3;

6. $S$  being drawn, the final estimate $X$ is given by equation~\ref{main_eq}. One starts again from step 2 for the next agent.

\subsection*{Comparison between theoretical and experimental results}

For all graphs, we ran $100000$ simulations so that the model predictions error bars are negligible.
Figure~\ref{fig1}B shows that the distribution of sensitivities to social influence $S$ obtained in the model (red curve) is similar by construction to the experimental one. Also by construction of the model (step 5. above), the cusp dependence of the social sensitivity with respect to $D=X_p-M$ is well reproduced by the model (Figure~\ref{fig4}A; red curve and empty symbols). We now address several non trivial predictions of the model.

\subsubsection*{Estimates after social influence}

Figure~\ref{fig1}A  (all values of $\rho$ aggregated) and SI Appendix, Fig.~S3 (for each $\rho$) compare favorably the distributions of estimates predicted by the model  with the experimental results (before and after social influence). Social influence leads to the sharpening of the distributions of estimates, and this effect increases as more information is provided to the group.

\subsubsection*{Impact of social information on collective performance}

\begin{figure}[ht]
\centering
	\includegraphics[width=.48\textwidth]{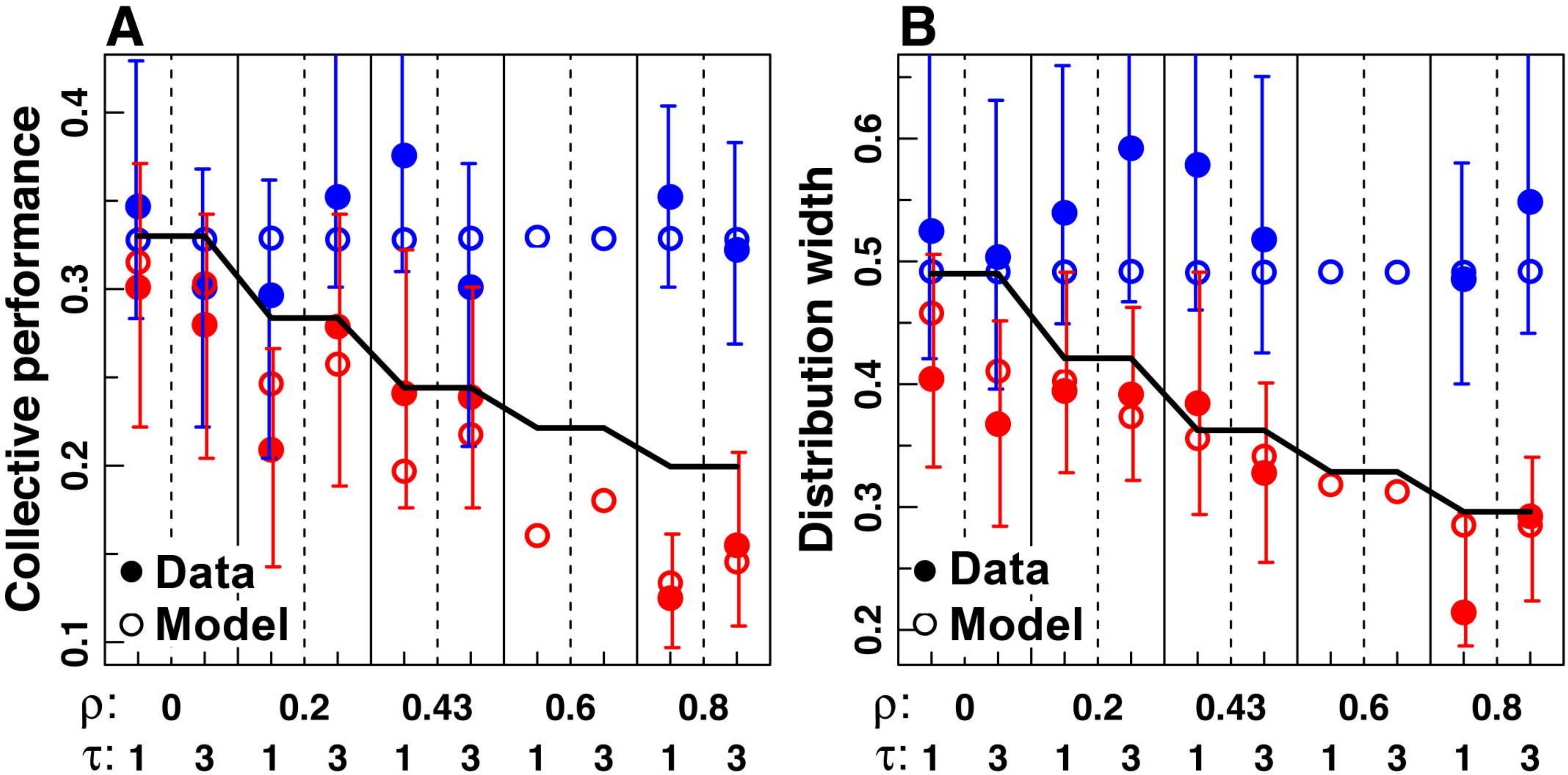}

\caption{Collective performance, defined as the absolute value of the median of estimates (A) and width of the distribution of estimates (B), for all  $(\rho,\tau)$, before (blue) and after (red) social influence. Both improve with $\rho$, as well as after social influence, except for the collective performance at $\rho = 0\,\%$. Full circles correspond to experimental data, while empty circles represent the predictions of the full model. The full black lines are the predictions of the simple solvable model presented in SI Appendix. For $\rho = 60\,\%$, only model predictions are available.}
 \label{fig2}
\end{figure}

Figure~\ref{fig2} shows the collective performance (precisely defined above) and the width of the distribution of estimates, for the different $\rho$ and $\tau$. The collective performance is $0$ when the distribution is centered on the true value, such that the closer it is to $0$, the better. As expected, when $\rho = 0\,\%$, no significant improvement is observed in the collective performance. Then, as $\rho$ increases, the center gets closer to the true value, and the width decreases accordingly, as also observed in the experiments  in Japan (see SI Appendix, Fig.~S9). Note that the experimental error bars (see SI Appendix for their computation) decrease after social influence, reflecting the decrease of the width of the estimate distribution after social influence and the driving of people's opinion by the virtual experts.
The collective performance and estimate distribution width predicted by the model (Figure~\ref{fig2}; empty circles) are in good agreement with those observed in the experiment.  The very small effect of $\tau$, only reliably observed in the model in Figure~\ref{fig2}A, is explained in the SI Appendix. As shown there, a simpler model where we neglect the dependence of $S$ with $D=X_p-M$ (Figure~\ref{fig4}A) can be analytically solved. It leads to fair predictions (full black lines on  Figure~\ref{fig2}), although it tends to underestimate the collective performance improvement  and does not capture the reduction of the distribution width already observed at $\rho=0\,\%$. This model guided us to design our experiments and its relative failure motivated us to investigate the phenomenon illustrated in Figure~\ref{fig4} and  included in the full model described above.

\subsubsection*{Impact of sensitivity to social influence on collective accuracy} \label{accuracy}

\begin{figure}[ht]
		\centering
		\includegraphics[width=.48\textwidth]{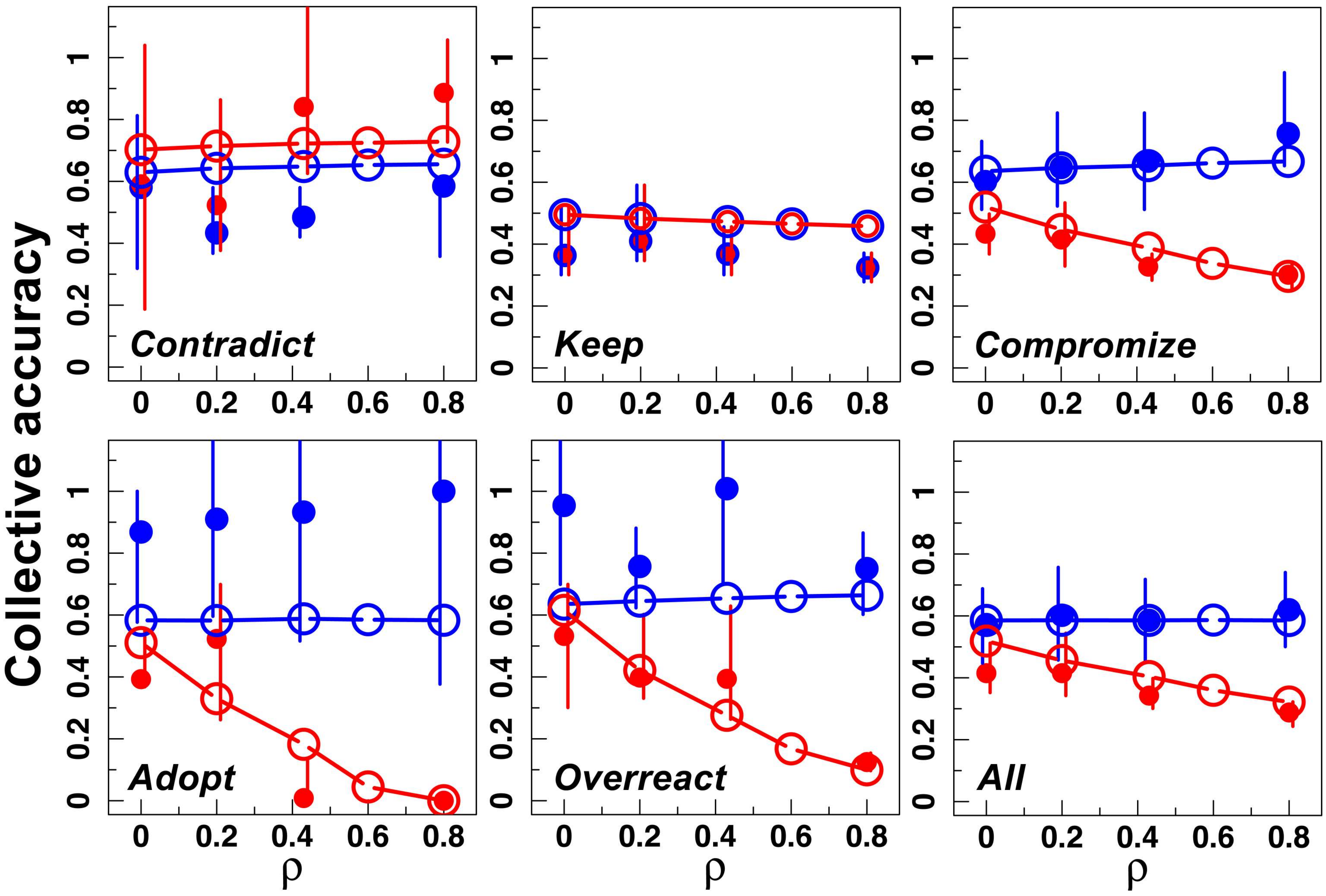}
		\caption{Collective accuracy (median distance to the truth of individual estimates) before (blue) and after (red) social influence against $\rho$, for the $5$ behavioral categories identified in Figure~\ref{fig1}B and for the whole group (All). Adopting leads to the sharpest improvement, and the best accuracy for $\rho \geq 40\,\%$.
Full circles correspond to experimental data, while empty circles represent the predictions of the model (including for $\rho=60$\,\%, a case not tested experimentally).}
	 \label{fig3}
\end{figure}

Figure~\ref{fig3} (see SI Appendix, Fig.~S11 for an alternative representation) shows the collective accuracy for the five categories of behavioral responses identified in Figure~\ref{fig1}B, and for the whole group, before and after social information.
Before social influence,  keeping  (Figure~\ref{fig3}B) leads to the best accuracy while  adopting and overreacting behaviors (Figure~\ref{fig3}D and E) are associated with the worst accuracy. However, as more reliable information is indirectly provided by the experts, and in particular for $\rho \geq 40\,\%$, adopting and overreacting lead to the best accuracy after social influence  \cite{kerckhove_modelling_2015,yaniv_receiving_2004}. The contradicting behavior (Figure~\ref{fig3}A) is the only one for which the accuracy is deteriorating after social influence. Finally, compromising (Figure~\ref{fig3}C) leads to a systematic improvement of the accuracy as the percentage of experts increases (better than keeping for $\rho \geq 40\,\%$), very similar to that of the whole group (Figure~\ref{fig3}F).
The collective accuracy for each behaviorial category is again fairly well predicted by the model (we discuss below the disagreement between model predictions and experimental data  in Figure~\ref{fig3}D, for the adopters before social influence).

The sensitivity to social influence and the collective accuracy are strongly related to confidence (see SI Appendix, Fig.~S10). The more confident the subjects, the less they tend to follow the group and the better is their accuracy, especially before social influence. This makes the link between confident \cite{madirolas_improving_2015}, informed \cite{faria_leadership_2010}, and successful \cite{king_is_2012} individuals: they are generally the same persons.
Yet,  individuals who are too confident (keeping behavior; arguably because they have an idea about the answer, hence their good accuracy before social influence), tend to discard others' opinion. Although it might sometimes work -- especially if no external information is provided $(\rho=0\,\%)$ -- they lose the opportunity to benefit from valuable information learned by others. Meanwhile, adopting and overreacting subjects have a poor confidence and accuracy before social influence (Figure~\ref{fig3}D and E), arguably because they do not know much about the question. Note that the model, not including any notion of confidence or heterogeneous prior knowledge, overestimates the quality of the accuracy before social influence for the adopting behavior. Yet, even at $\rho = 0\,\%$, adopting subjects perform about as well as the other categories after social influence. In fact, if enough information is provided ($\rho =80\,\%$), they are even able to reach almost perfect collective accuracy. Similar results have been found in the Japan experiment, as shown on SI Appendix, Fig.~S12. SI Appendix, Fig.~S13, S14, and S15 show  similar graphs for the collective performance in France and Japan.

\subsubsection*{Predicting the effect of incorrect information given to the human group by virtual agents}

We used the model to investigate the influence on the group performance of the quality and quantity of information delivered to the group, \emph{i.e.} the value $V$ of the answer provided by the percentage $\rho$ of virtual agents. In our experiments, the group was provided with the (log-transform of the) true value $V = 0$ (the agents were ``experts''). We expect a deterioration of the collective performance and accuracy as $V$ moves too far away from 0, and as a greater amount of incorrect information is delivered to the group (by increasing $\rho$).  The optimum collective accuracy is reached for $V$\emph{ strictly positive}, whatever the value of $\rho>0$ (SI Appendix, Fig.~S16), as also predicted by our simple analytical model. Hence, incorrect information can be beneficial to the group: providing the group with overestimated values can counterbalance the human cognitive bias to underestimate quantities \cite{indow_scaling_1977}.

\section*{Discussion}

Quantifying how social information affects individual estimations and opinions is a crucial step to understanding and modeling the dynamics of collective choices or opinion formation \cite{ball_why_2012}. Here, we have measured and modeled the impact of social information at individual and collective scales in estimation tasks with low demonstrability. By controlling the quantity and quality of information delivered to the subjects, unbeknownst to them, we have been able to precisely quantify the impact of social influence on group performance. We also tested and confirmed the cross-cultural generality of our results by conducting experiments in France and Japan.

We showed and justified that when individuals have poor prior knowledge about the questions, the distribution of their log-transformed estimates is close to a Cauchy distribution. The distribution of the sensitivity to social influence $S$ is bell-shaped (contradict, compromise, overreact), with two additional peaks exactly at $S=0$ (keep) and $S=1$ (adopt), which lead to the definition of robust social traits, as checked by further observing the subjects inclined to follow these behaviors. When subjects have little prior knowledge, we found that their sensitivity to social influence increases (linear cusp) with the difference between their estimate and that of the group, at variance with what was found in \cite{yaniv_receiving_2004}, for questions where subjects had a high prior knowledge.

We used these experimental observation to build and calibrate a model that quantitatively predicts the sharpening of the distribution of individual estimates and the improvement in collective performance and accuracy, as the amount of good information provided to the group increases. This model could be directly applied or straightforwardly adapted to similar situations where humans have to integrate  information from other people or external sources.

We studied the impact of virtual experts on the group performance, a methodology allowing to rigorously control the quantity ($\rho$) and quality ($V$) of the information provided to a group with little prior knowledge. These virtual experts can be seen either as an external source of information accessible to individuals (\emph{e.g.}, Internet, social networks, medias...), or as a  very cohesive (all having the same opinion $V$) and over-confident (all having $S=0$) subgroup of the population, as can happen with ‘‘groupthink’’ \cite{janis_groupthink:_1982}. When these experts provide reliable information to the group, a systematic improvement in collective performance and accuracy is obtained experimentally and is quantitatively reproduced by our model.
Moreover, if the experts are not too numerous and the information they give is slightly above the true value, the model predicts that social influence can help the group perform even better than when the truth is provided, as this incorrect information compensates for the human cognitive bias to underestimate quantities.

We also showed that the sensitivity to social influence is strongly related to confidence and accuracy: the most confident subjects are generally the best performers, and tend to weight the opinion of others less. When the group has access to more reliable information, this behavior becomes detrimental to individual and collective accuracy, as too confident individuals lose the opportunity to benefit from this information.

Overall, we showed that individuals, even when they have very little prior knowledge about a quantity to estimate, are able to use information from their peers or from the environment, to collectively improve the group performance, as long as this information is not highly misleading. Ultimately, getting a better understanding of these influential processes opens new perspectives to develop information systems aimed at enhancing cooperation and collaboration in human groups, thus helping crowds become smarter \cite{helbing_globally_2013,xhafa_inter-cooperative_2014}.

Future research will have to focus on the experimental validation of our theoretical predictions when providing incorrect information to the group, with the intriguing possibility to actually improve its performance. It would also be interesting to study the impact on the group performance of the number of estimates given as social information (instead of only their mean), and of revealing the confidence and/or reputation of those who share these estimates.


\acknow{This work was supported by the project ANR-11-IDEX-0002-02 – Transversalité - MuSE, by a grant from the CNRS Mission for Interdisciplinarity (project SmartCrowd, AMI S2C3), and by ‘‘Programme Investissements d'Avenir’’, under the program ANR-11-IDEX-0002-02, reference ANR-10-LABX-0037-NEXT. B.J. was supported by a doctoral fellowship from the CNRS, R.E. was supported by Marie Curie core/programme grant funding, grant Number 655235 - SmartMass. T.K. was supported by JSPS KAKENHI Grant Numbers JP16H06324 and JP25118004. We are grateful to Pr. Ofer Tchernichovski for his valuable comments.}
\showacknow

\pnasbreak

\bibliography{biblio}

\end{document}